\documentclass[aps,prl,twocolumn,amsmath,amssymb,footinbib,showpacs,longbibliography,superscriptaddress]{revtex4-1}

\newcommand{\Jnature}{Nature (London)}
\newcommand{\Jnatphys}{Nat. Phys.}
\newcommand{\Jnatcomm}{Nat. Comm.}

\newcommand{\Jprl}{Phys. Rev. Lett.}

\newcommand{\Jpra}{Phys. Rev. A}
\newcommand{\Jprb}{Phys. Rev. B}

\newcommand{\Jpre}{Phys. Rev. E}
\newcommand{\Jrmp}{Rev. Mod. Phys.}

\newcommand{\Jepl}{Europhys. Lett.}

\newcommand{\Jepjd}{Eur. Phys. J. D}

\newcommand{\JphyslettA}{J. Phys. Lett. A}

\newcommand{\JRepProgPhys}{Rep. Prog. Phys.}

\newcommand{\Jadvphys}{Adv. Phys.}

\newcommand{\Jannphys}{Ann. Phys. (NY)}

\newcommand{\JAnnualRevCondMat}{Annual Rev. Cond. Mat. Phys.}

\newcommand{\JMathPhys}{J. Math. Phys.}

\usepackage[english]{babel}
\usepackage{latexsym}
\usepackage{graphics}
\usepackage{subfigure}
\usepackage{epsfig}
\usepackage{color}
\usepackage{hyperref}
\usepackage{braket} 
\usepackage[T1]{fontenc}
\usepackage[latin9]{inputenc}
\usepackage{amstext}
\usepackage{amssymb}
\usepackage{graphicx}
\usepackage{esint}
\usepackage{braket}
\usepackage{babel}
\usepackage{amsmath}

\hypersetup{
colorlinks=true,
citecolor=blue,
linkcolor=red,
urlcolor=black
}


\newcommand{\e}{\textrm{e}}
\newcommand{\ie}{i.e.}

\newcommand{\IPR}{\textrm{IPR}}
\newcommand{\Er}{E_{\textrm{r}}}
\newcommand{\Ec}{E_{\textrm{c}}}
\newcommand{\Vc}{V_{\textrm{c}}}
\newcommand{\Tm}{T_{\textrm{m}}}

\newcommand{\kB}{k_{\textrm{\tiny {B}}}}

\newcommand{\fED}{f_{\textrm{\tiny {FD}}}}

\renewcommand{\DH}{D_\textrm{\tiny H}}
\newcommand{\NB}{N_\textrm{\tiny B}}

\newcommand{\aoneD}{a_{\textrm{\tiny 1D}}}

\newcommand{\fs}{f_{\textrm{s}}}
\newcommand{\ups}{\Upsilon_\textrm{s}}
\newcommand{\aOneD}{a_{\textrm{\tiny 1D}}}

\newcommand{\lettersection}[1]{\paragraph*{#1.---}}

\begin{document}

\title{
Lieb-Liniger Bosons in a Shallow Quasiperiodic Potential: Bose Glass Phase and Fractal Mott Lobes
}

\author{Hepeng Yao}
\affiliation{CPHT, CNRS, Institut Polytechnique de Paris, Route de Saclay 91128 Palaiseau, France}

\author{Thierry Giamarchi}
\affiliation{Department of Quantum Matter Physics, University of Geneva, 24 Quai Ernest-Ansermet, CH-1211 Geneva, Switzerland}

\author{Laurent Sanchez-Palencia}
\affiliation{CPHT, CNRS, Institut Polytechnique de Paris, Route de Saclay 91128 Palaiseau, France}

\date{\today}

\begin{abstract}
The emergence of a compressible insulator phase, known as the Bose glass, is characteristic of the interplay of interactions and disorder in correlated Bose fluids. While widely studied in tight-binding models, its observation remains elusive owing to stringent temperature effects. Here we show that this issue may be overcome by using Lieb-Liniger bosons in shallow quasiperiodic potentials. A Bose glass, surrounded by superfluid and Mott phases, is found above a critical potential and for finite interactions. At finite temperature, we show that the melting of the Mott lobes is characteristic of a fractal structure and find that the Bose glass is robust against thermal fluctuations up to temperatures accessible in quantum gases. Our results raise questions about the universality of the Bose glass transition in such shallow quasiperiodic potentials.
\end{abstract}

\maketitle


The interplay of interactions and disorder in quantum fluids is at the origin of many intriguing phenomena,
including many-body localization~\cite{basko2006,oganesyan2007,nandkishore2015,altman2015,abanin2019},
collective Anderson localization~\cite{gurarie2002,gurarie2003,gurarie2008,bilas2006,lugan2007b,lugan2011,lellouch2014,lellouch2015},
and the emergence of new quantum phases.
For instance, a compressible insulator, known as the Bose glass (BG)~\cite{giamarchi1987,giamarchi1988,fisher1989,krauth1991,rapsch1999},
may be stabilized against the superfluid (SF)
and, in lattice models, against the Mott insulator (MI).
One-dimensional (1D) systems are particularly fascinating
for the SF may be destabilized by arbitrary weak perturbations,
an example of which is the pinning transition in periodic potentials~\cite{haldane1980,haldane1981,cazalilla2011,haller2010,boeris2016}.
Similarly, above an interaction threshold, the BG transition can be induced by arbitrary weak disorder~\cite{giamarchi1987,giamarchi1988}.
The phase diagram of 1D disordered bosons has been extensively studied
and is now well characterized theoretically~\cite{scalettar1991,krauth1991,rapsch1999,lugan2007a,fontanesi2009,vosk2012}.
The experimental observation of the BG phase remains, however, elusive~\cite{fallani2007,lahini2008,pasienski2010,deissler2010,gadway2011,yu2012},
despite recent progress using ultracold atoms in quasiperiodic potentials~\cite{derrico2014,gori2016}. 

Controlled quasiperiodic potentials, as realized in ultracold atom~\cite{lewenstein2007,modugno2010,lsp2010} and photonic~\cite{lahini2009,verbin2015,tanese2014,barboux2017} systems, have long been recognized as a promising alternative to observe the BG phase.
So far, however, this problem has been considered only in the tight-binding limit, known as the Aubry-Andr\'e model~\cite{damski2003,roth2003,roscilde2008,deng2008,roux2008}.
It sets the energy scale to the tunneling energy, which is exponentially small in the main lattice amplitude and of the order of magnitude of the temperature in typical experiments. The phase coherence is then strongly reduced, which significantly alters the phase diagram. Although such systems give some evidence of a Bose glass phase~\cite{derrico2014,gori2016}, they require a heavy heuristic analysis of the data to factor out the very important effects of the temperature. 

Here, we propose to overcome this issue by using shallow quasiperiodic potentials.
The energy scale would then be the recoil energy, which is much larger than typical temperatures in ultracold-atom experiments~\cite{lewenstein2007,bloch2008}.
This, however, raises the fundamental question of whether a BG phase can be stabilized in this regime:
In the hard-core limit, interacting bosons map onto free fermions~\cite{girardeau1960}.
A band of localized (resp.\ extended) single particles then maps onto the BG (resp.\ SF) phase while a band gap maps onto the MI phase. In the shallow bichromatic lattice, however, it has been shown that band gaps, \ie\ MI phases, are dense~\cite{yao2019} and the BG would thus be singular.
On the other hand, decreasing the interactions down to the mean field regime favors the SF phase~\cite{giamarchi1987,giamarchi1988,lsp2006,lugan2007a}.
Hence, a BG can only be stabilized, if at all, for intermediate interactions.

\begin{figure*}[t!]
\centering
        \includegraphics[width = 1.9\columnwidth]{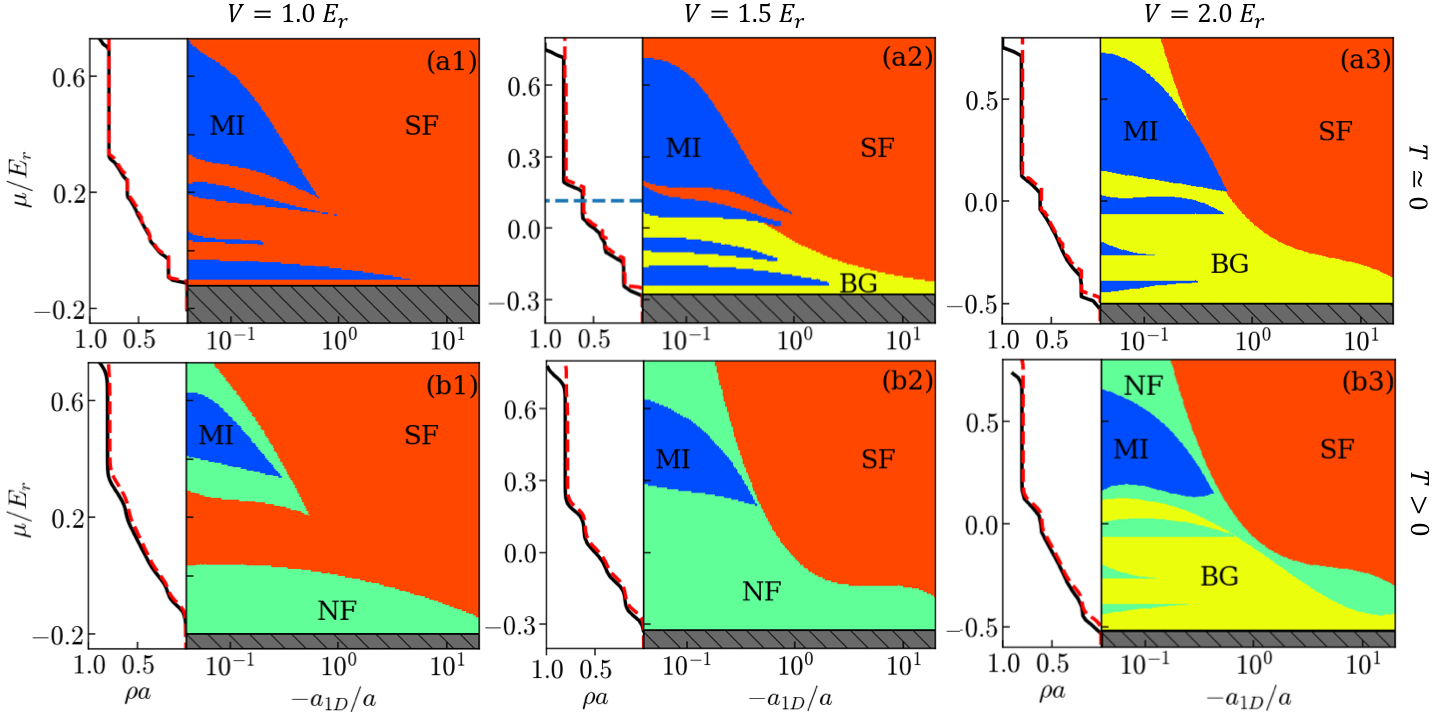}
\caption{\label{fig:phase}
Phase diagrams of Lieb-Liniger bosons in a shallow quasiperiodic potential for increasing values of the potential ($V=\Er<\Vc$, $V=1.5\Er\gtrsim \Vc$, and $V=2\Er>\Vc$).
Upper row:~Quantum phase diagrams from QMC calculations at a vanishingly small temperature [$\kB T=10^{-3}\Er$ for (a1) and $\kB T=2\times 10^{-3}\Er$ for (a2) and (a3)].
Lower row:~Counterpart of the upper row at the finite temperature $\kB T=0.015\Er$.
On the left of each panel, we show the equation of state $\rho(\mu)$ at strong repulsive interactions, $-\aoneD/a=0.05$ (solid black line) together with that of free fermions at the corresponding temperatures (dashed red line).
Note that the smallest band gaps are smoothed out by the finite temperatures~\cite{note:SupplMat}.
The dotted blue line in (a2) shows the single-particle ME at $V=1.5\Er$, $\Ec \simeq 0.115\Er$.
}
 \end{figure*}

We tackle this issue using exact quantum Monte Carlo (QMC) calculations.
We compute the exact phase diagram of interacting bosons in a shallow 1D bichromatic lattice. Our results are summarized in Fig.~\ref{fig:phase}.
Our main finding is that, at zero temperature, a significant BG phase can be stabilized above a critical quasiperiodic amplitude and for intermediate interaction strengths (see upper row in Fig.~\ref{fig:phase}).
Further, we study the melting of the quantum phases at finite temperature. We show that their main features are robust against thermal fluctuations up to temperatures accessible to experiments, in spite of the growth of a normal fluid (NF) regime (see lower row on Fig.~\ref{fig:phase}).
Moreover, while the SF and BG phases progressively cross over to the NF regime, we find that the MI phase shows a transient anomalous temperature-induced enhancement of coherence.
We show that the melting of the MI phase presents a characteristic algebraic temperature dependence,
which we relate to the fractal structure of the MI lobes.

\lettersection{Model and approach}

The system we consider is a Lieb-Liniger gas, \ie\ a 1D $N$-boson gas with repulsive contact interactions, subjected to a quasiperiodic potential $V(x)$. It is governed by the Hamiltonian
\begin{equation}\label{eq:Hamiltonian}
\mathcal{H} =\sum_{1 \leq j \leq N} \Big[-\frac{\hbar^2}{2m}\frac{\partial^2}{\partial x_j^2}+V(x_j)\Big]+g\sum_{j<\ell}\delta(x_j-x_\ell),
\end{equation}
where $m$ is the particle mass, $x$ the space coordinate, $g=-2\hbar^2/m\aoneD$ the coupling constant, and $\aoneD<0$ the 1D scattering length (corresponding to repulsive interactions~\cite{olshanii1998}).
The quasiperiodic potential is bichromatic with equal amplitudes, \ie\,
\begin{equation}\label{eq:QPpotential}
V(x) = \frac{V}{2} \big[ \cos \left(2k_1 x\right) + \cos \left(2k_2 x + \varphi\right) \big],
\end{equation}
with incommensurate spatial frequencies $k_1$ and $k_2$, and we used $\varphi=0.2$.
Qualitatively similar results are expected for imbalanced potentials and different incommensurate ratios $r=k_2/k_1$, although the value of the critical localization potential may vary~\cite{biddle2009,yao2019}.
In this respect, the balanced case is nearly optimal~\cite{yao2019} and we choose $r \simeq 0.807$, close to the experiments of Refs.~\cite{derrico2014,gori2016,note:size}.
In the following, we use the spatial period of the first lattice, $a=\pi/k_1$, and the corresponding recoil energy, $\Er=\hbar^2 k_1^2/2m$, as the space and energy units, respectively.
Using these units, typical interaction and temperature ranges in recent experiments are $-\aOneD/a\simeq 0.06-20$ and $\kB T/\Er\simeq 0.015-0.15$~\cite{haller2010,derrico2014,meinert2015,boeris2016}.

Let us start with the single-particle problem, which is relevant to the hard-core limit.
The localization and spectral properties of single particles in the shallow bichromatic lattice Eq.~(\ref{eq:QPpotential}) have been studied previously~\cite{biddle2009,biddle2010,szabo2018,yao2019}.
Below some critical amplitude $\Vc \sim \Er$, all the states are extended,
while above $\Vc$, a band of localized states appears at low energy, up to an energy mobility edge (ME) $\Ec$.
The existence of a finite $\Ec$ distinguishes the shallow lattice model from the celebrated Aubry-Andr\'e tight-binding model, where it is absent.
We determine the single-particle eigenstates using exact diagonalization and the
critical amplitude is precisely found from the finite-size scaling analysis of their inverse participation ratio (IPR). We then find $\Vc/\Er \simeq 1.38\pm0.01$~\cite{note:SupplMat}.

\lettersection{Phase diagrams for interacting bosons}
We now turn to the interacting Lieb-Liniger gas.
At zero temperature, we expect three possible phases:
the MI (incompressible insulator), the SF (compressible superfluid), and the BG (compressible insulator).
They are identified through the values of
the compressibility $\kappa$ and the superfluid fraction $\fs$.
We have also checked that the one-body correlation function $g_1(x)=\int \frac{dx^{\prime}}{L}\langle \Psi(x^{\prime}+x)^\dagger\Psi(x^{\prime})\rangle$
decays exponentially in the insulating phases (MI and BG) and algebraically in the SF phase (see below).
All these quantities are found using quantum Monte Carlo calculations in continuous space within the grand-canonical ensemble (temperature $T$ and particle chemical potential $\mu$)~\cite{note:SupplMat}.

The upper row in Fig.~\ref{fig:phase} shows the quantum phase diagrams
versus the inverse interaction strength and the chemical potential,
for increasing amplitudes of the quasiperiodic potential.
They are found from QMC calculations of $\kappa$ and $\fs$ at a vanishingly small temperature~\cite{note:SupplMat}.
In practice, we have used $\kB T \sim 0.001-0.002 \Er$, where $\kB$ is the Boltzmann constant, and we have checked that there is no sizable temperature dependence at a lower temperature.
For $V<\Vc$, no localization is expected and we only find SF and MI phases; see Fig.~\ref{fig:phase}(a1).
The SF dominates at large chemical potentials and weak interactions.
Strong enough interactions destabilize the SF phase and Mott lobes open, with fractional occupation numbers ($\rho a=r$, $2r-1$, $2-2r$, $1-r$ from top to bottom).
The number of lobes increases with the interaction strength and eventually become dense in the hard-core limit (see below).
For $V>\Vc$ and a finite interaction, a BG phase
develops in between the MI lobes up to the single-particle ME at $\mu=\Ec$; see Fig.~\ref{fig:phase}(a2).
There, the SF fraction is strictly zero and the compressibility has a sizable, nonzero value, within QMC accuracy.
When the quasiperiodic amplitude $V$ increases, the BG phase extends at the expense of both the MI and SF phases; see Fig.~\ref{fig:phase}(a3).

The lower row in Fig.~\ref{fig:phase} shows the counterpart of the previous diagrams at the finite temperature $T=0.015\Er/\kB$, corresponding to the minimal temperature in Ref.~\cite{meinert2015}.
While quantum phases may be destroyed by arbitrarily small thermal fluctuations,
the finite-size systems we consider ($L=83 a$, of the order of typical sizes in experiments~\cite{fallani2007,deissler2010,derrico2014}) retain characteristic properties,
reminiscent of the zero-temperature phases.
The SF, MI, and BG \textit{regimes} shown in Figs.~\ref{fig:phase}(b1)-(b3) are identified accordingly.
While the former two are easily identified, special care should be taken for the BG, which cannot be distinguished from the normal fluid via $\kappa$ and $\fs$, since both are compressible insulators.
A key difference, however, is that correlations are suppressed by the disorder in the BG and by thermal fluctuations in the NF.
To identify the BG regime, we thus further require that the suppression of correlations is dominated by the disorder, \ie\ the correlation length is nearly independent of the temperature.
The QMC results show that the NF develops at low density and strong interactions; see Fig.~\ref{fig:phase}(b1).
For a moderate quasiperiodic amplitude, it takes over the BG, which is completely destroyed; see Fig.~\ref{fig:phase}(b2).
For a strong enough quasiperiodic potential, however, the BG is robust against thermal fluctuations and competes favorably with the NF regime; see Fig.~\ref{fig:phase}(b3).
We hence find a sizable BG regime, which should thus be observable at temperatures accessible to  current experiments using 1D quantum gases.

\begin{figure}[t!]
         \centering
         \includegraphics[width = 1.0\columnwidth]{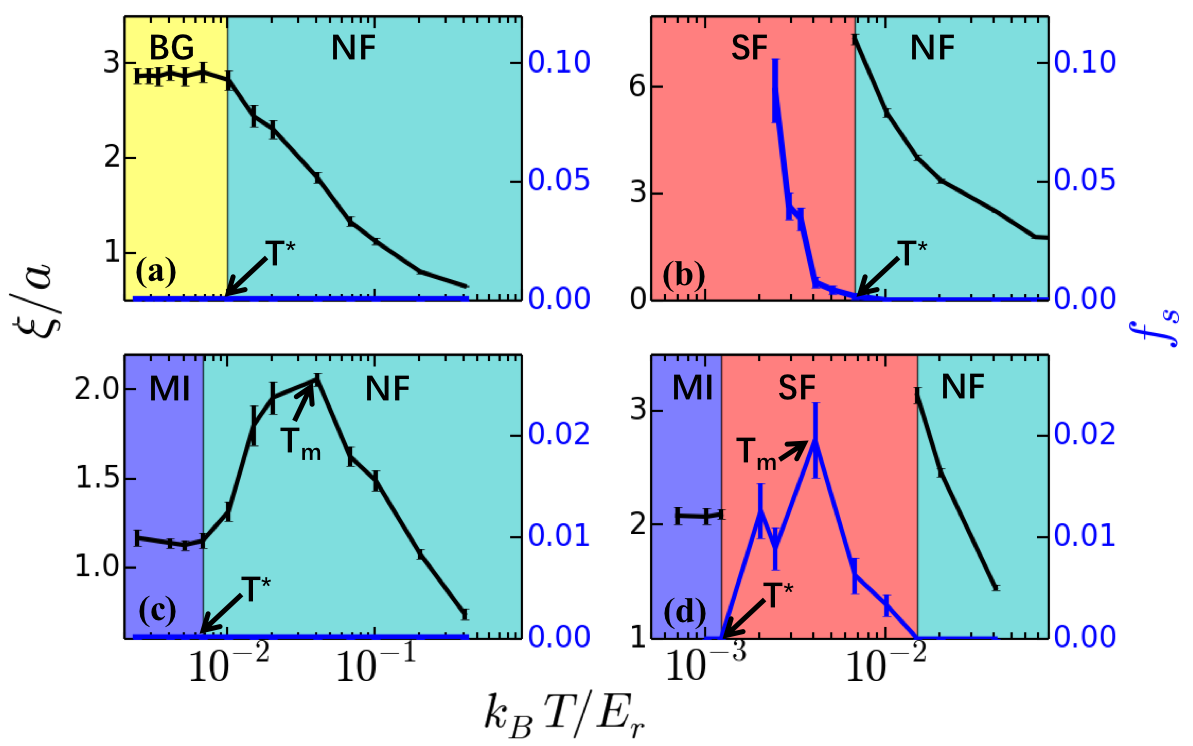}
\caption{\label{fig:melting}
Temperature-induced melting of the quantum phases.
The various panels show the coherence length $\xi$ (black lines) and the superfluid fraction $\fs$ (blue lines) as a function of temperature for four typical cases.
(a)~BG to NF crossover ($V=2\Er$, $\mu=-0.28\Er$, and $-\aOneD/a=4.0$),
(b)~SF to NF crossover ($V=2\Er$, $\mu=0.53\Er$, and $-\aOneD/a=0.2$),
(c)~MI to NF crossover ($V=2\Er$, $\mu=0.47\Er$,and $-\aOneD/a=0.2$),
(d)~MI to NF, via SF, crossover ($V=\Er$, $\mu=0.2\Er$, and $-\aOneD/a=0.05$).
}
 \end{figure}

\lettersection{Melting of the quantum phases}
We now turn to the quantitative study of the temperature effects.
We compute the one-body correlation function and fit it to $g_1(x) \sim \exp\left(-\vert x\vert/\xi\right)$, where $\xi$ is the correlation length~\cite{note:SupplMat}.
The typical behavior of $\xi(T)$ when increasing the temperature $T$ from a point in the BG phase is displayed in Fig.~\ref{fig:melting}(a) (black line).
It shows a plateau at low temperature, which is identified as the BG regime.
Above some melting temperature $T^*$, the thermal fluctuations suppress phase coherence and $\xi$ decreases with $T$, as expected for a NF.
In both the BG and NF regimes, superfluidity is absent and we consistently find $\fs=0$, also shown in Fig.~\ref{fig:melting}(a) (blue line).

Consider now increasing the temperature from a point in the SF phase at $T=0$; see Fig.~\ref{fig:melting}(b). For low enough $T$, we find a finite SF fraction $\fs$, which, however, strongly decreases with $T$.
The sharp decrease of $\fs$ allows us to identify a rather well-defined temperature $T^*$ beyond which we find a NF regime, characterized by a vanishingly small $\fs$.
We checked that, consistently, the correlation function turns at $T^*$
from a characteristic algebraic to exponential decay over the full system of length $L=83 a$~\cite{note:SupplMat}.

Consider finally increasing the temperature from a point in a MI lobe at $T=0$, see two typical cases in Figs.~\ref{fig:melting}(c) and ~\ref{fig:melting}(d).
As expected, below the melting temperature $T^*$, the correlation length shows a plateau, identified as the MI regime.
Quite counterintuitively, however, we find that above $T^*$ the phase coherence is enhanced by thermal fluctuations, up to some temperature $\Tm$, beyond which it is finally suppressed.
This anomalous behavior is signaled by the nonmonotony of the correlation length $\xi(T)$; see Fig.~\ref{fig:melting}(c).
In some cases, it is strong enough to induce a finite superfluid fraction $\fs$, and correspondingly an algebraic correlation function in our finite-size system~\cite{note:SupplMat}, see Fig.~\ref{fig:melting}(d). This is typically the case when the MI lobe is surrounded by a SF phase at $T=0$.
We interprete this behavior from the competition of two effects.
On the one hand, a finite but small temperature permits the formation of particle-hole pair excitations, which are extended and support phase coherence. This effect, which is often negligible in strong lattices, is enhanced in shallow lattices owing to the smallness of the Mott gaps, particularly in the quasiperiodic lattice where Mott lobes with fractional fillings appear~\cite{roscilde2008,yao2019}. This favors the onset of a finite-range coherence at finite temperature. On the other hand, when the temperature increases, a larger number of extended pairs, which are mutually incoherent, is created. This suppresses phase coherence on a smaller and smaller length scale, hence competing with the former process and leading to the nonmonotonic temperature dependence of the coherence length.

\begin{figure}[t!]
         \centering
         \includegraphics[width = 1.0\columnwidth]{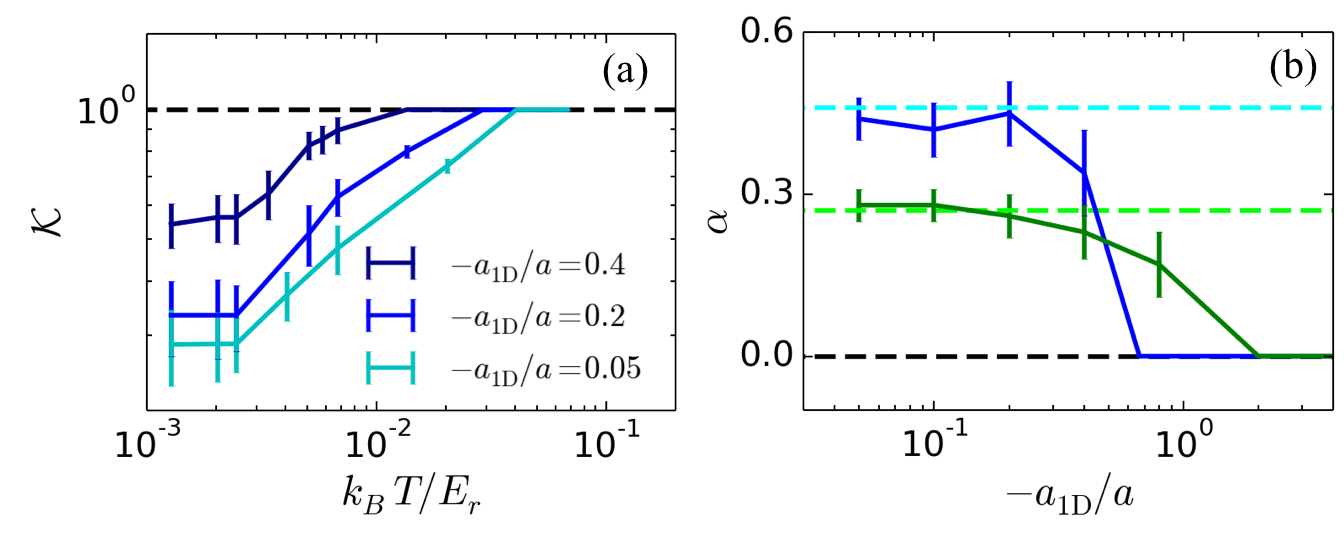}
\caption{\label{fig:fractalMIlobes}
Melting of the Mott phase.
(a)~Compressible phase fraction versus temperature for $V=1.5\Er$, $\mu/\Er \in[-0.4, 0.8]$, and various interaction strengths.
(b)~Exponent $\alpha$ versus the interaction strength for $V=\Er$ (green solid line) and $V=1.5\Er$ (blue).
The colored dashed lines indicate the corresponding values of $1-\DH$.
}
 \end{figure}

\lettersection{Fractal Mott lobes}
The melting of a Mott lobe of gap $\Delta$ is expected at a temperature $T \propto \Delta/\kB$~\cite{gerbier2007}.
In the quasiperiodic lattice, however, there is no typical gap, owing to the fractal structure of the Mott lobes,
inherited from that of the single-particle spectrum~\cite{roscilde2008,roux2008,yao2019}.
To get further insight into the melting of the MI lobes, consider the \textit{compressible phase fraction}, \ie, the complementary of the fraction of MI lobes,
\begin{equation}\label{eq:fraction}
\mathcal{K}=\lim_{q\rightarrow0_+}\int_{\mu_1}^{\mu_2}\frac{d \mu}{\mu_2-\mu_1}\  \big[\kappa(\mu)\big]^q,
\end{equation}
in the chemical potential range $[\mu_1,\mu_2]$.
The behavior of $\mathcal{K}$ versus temperature is shown in Fig.~\ref{fig:fractalMIlobes}(a) for various interaction strengths. Below the  melting temperature of the smallest MI lobes, $T_1^{\star}$, $\mathcal{K}$ is insensitive to thermal fluctuations and we correspondingly find $\mathcal{K}=\textrm{constant}$. Above that of the largest lobe, $T_2^{\star}$, all MI lobes are melt and $\mathcal{K}=1$~\cite{note:Tstar}. In the intermediate regime, $T_1^{\star}\lesssim T\lesssim T_2^{\star}$, we find the algebraic scaling $\mathcal{K}\sim T^{\alpha}$, where the exponent $\alpha$ depends on both the interaction strength and the quasiperiodic amplitude $V$, see Fig.~\ref{fig:fractalMIlobes}(b). This behavior is reminiscent of the fractal structure of the MI lobes.

To understand this, consider the Tonks-Girardeau limit, $\aOneD\rightarrow0$, where the Lieb-Liniger gas may be mapped onto free fermions~\cite{girardeau1960}. The particle density then reads as $\rho(\mu)\simeq \frac{1}{L}\sum_j \fED(E_j-\mu)$, where $\fED(E)=1/(e^{E/\kB T}+1)$ is the Fermi-Dirac distribution and $E_j$ is the $j$ th eigenenergy of the single-particle Hamiltonian. This picture provides a very good approximation of our QMC results at  large interaction, irrespective of $T$ and $V$; see left-hand panels of each plot in Fig.~\ref{fig:phase}.
The compressibility thus reads as
$\kappa(\mu)\simeq\frac{-1}{L} \sum_j \fED^\prime (E_j-\mu)$.
Since $\fED^\prime=\partial \fED/\partial E$ is a peaked function of typical width $\kB T$ around the origin, we find
\begin{equation}\label{eq:kappa-FD}
\kappa(\mu) \sim n_{\epsilon=\kB T}(\mu),
\end{equation}
where $n_{\epsilon}(E)$ is the integrated density of states per unit length of the free Hamiltonian in the energy range $[E-\epsilon/2,E+\epsilon/2]$. Hence, the compressibility maps onto the integrated density of states, $\kB T$ onto the energy resolution, and, up to the factor $\kB T / (\mu_2-\mu_1)$, the compressible phase fraction onto the spectral box-counting number $N_\textrm{B}(\epsilon)$ introduced in Ref.~\cite{yao2019}.
We then find
\begin{equation}\label{eq:box-counting}
\mathcal{K} \sim \frac{\kB T}{\mu_2-\mu_1} N_\textrm{B}(\epsilon=\kB T) \sim T^{1-\DH},
\end{equation}
where $\DH$ is the Hausdorff dimension of the free spectrum~\cite{note:SupplMat},
and we recover the algebraic temperature dependence $\mathcal{K} \sim T^\alpha$, with $\alpha=1-\DH$.

To validate this picture, we have computed the exponent $\alpha$ by fitting curves as in Fig.~\ref{fig:fractalMIlobes}(a) as a function of the interaction strength. The results are shown in Fig.~\ref{fig:fractalMIlobes}(b) for two values of the quasiperiodic amplitude (colored solid lines). As expected, we find $\alpha \rightarrow 1-\DH$ (colored dashed lines) in the Tonks-Girardeau limit, $\aOneD\rightarrow0$. When the interaction strength decreases, the fermionization picture breaks down. The exponent $\alpha$ then decreases and vanishes when the last MI lobe shrinks.

Moreover, our results show that the compressible BG fraction is suppressed at low temperature (since $\alpha >0$) and strong interactions [see Fig.~\ref{fig:fractalMIlobes}(a)]. This is consistent with the expected singularity of the BG phase in the hard-core limit, where the MI lobes become dense.

\lettersection{Conclusion}
In summary, we have computed the quantum phase diagram of Lieb-Liniger bosons in a shallow quasiperiodic potential. Our main result is that a BG phase emerges above a critical potential and for finite interactions, surrounded by SF and MI phases.
We have also studied finite-temperature effects.
We have shown that the melting of the MI lobes is characteristic of their fractal structure and found regimes where the BG phase is robust against thermal fluctuations up to a range accessible to experiments. This paves the way to the direct observation of the still elusive BG phase, as well as the fractality of the MI lobes, in ultracold quantum gases. 

More precisely, the temperature $T=0.015\Er/\kB$ used in Fig.~\ref{fig:phase} corresponds to $T \simeq 1.5$nK for $^{133}$Cs ultracold atoms, which is about the minimal temperature achieved in Ref.~\cite{meinert2015}.
Further, we have checked that a sizable BG regime is still observable at higher temperatures, for instance $T=0.1\Er/\kB$~\cite{note:SupplMat}, which is higher than the temperatures reported in Refs.~\cite{meinert2015,derrico2014}.
We propose to characterize the phase diagram using the one-body correlation function,
as obtained from Fourier transforms of time-of-flight images in ultracold atoms~\cite{derrico2014,gori2016}. Discrimination of algebraic and exponential decays could benefit from box-shaped potentials~\cite{meyrath2005,gaunt2013,chomaz2015}. Our results indicate that the variation of the correlation length $\xi(T)$ with the temperature characterizes the various regimes; see Fig.~\ref{fig:melting}.

Further, our work questions the universality of the BG transition found here.
In contrast to truly disordered~\cite{giamarchi1987,giamarchi1988} or Fibonacci~\cite{vidal1999,vidal2001} potentials, the shallow bichromatic lattice contains only two spatial frequencies of finite amplitudes.
Hence, the emergence of a BG requires the growth of a dense set of density harmonics within the renormalization group flow, which may significantly affect the value of the critical Luttinger parameter.

\begin{acknowledgments}
This research was supported by the
European Commission FET-Proactive QUIC (H2020 Grant No.~641122),
the Paris region DIM-SIRTEQ,
and the Swiss National Science Foundation under Division~II.
This work was performed using HPC resources from GENCI-CINES (Grant No. 2018-A0050510300).
Numerical calculations make use of the ALPS scheduler library and statistical analysis tools~\cite{troyer1998,ALPS2007,ALPS2011}. 
We thank the CPHT computer team for valuable support.
\end{acknowledgments}

 \renewcommand{\theequation}{S\arabic{equation}}
 \setcounter{equation}{0}
 \renewcommand{\thefigure}{S\arabic{figure}}
 \setcounter{figure}{0}
 \renewcommand{\thesection}{S\arabic{section}}
 \setcounter{section}{0}
 \onecolumngrid  
     
 
 \newpage

{\center \bf \large Supplemental Material for \\}
{\center \bf \large Lieb-Liniger Bosons in a Shallow Quasiperiodic Potential: Bose Glass Phase and Fractal Mott Lobes \\ \vspace*{1.cm}
}

In this supplemental material, we provide details about the localization and fractal properties of single-particle in the shallow quasiperiodic potential~(Sec.~\ref{sec:OneBody}), the quantum Monte Carlo data for the many-body Bose gas~(Sec.~\ref{sec:ManyBody}), as well as the discussion about the experimental realization~(Sec.~\ref{sec:exp}).

\section{Localization and fractal properties of single particles in the shallow bichromatic potential}
\label{sec:OneBody}

We determine the single-particle properties of the shallow quasiperiodic potential [Eq.~(2) of the main paper] similarly as in Ref.~\cite{yao2019}, but using here the ratio $r \simeq 0.807$.
We discuss the localization critical potential $\Vc$ and the energy mobility edge (ME) $\Ec$ in Sec.~\ref{app:VcEc},
and the spectral fractal properties in Sec.~\ref{app:fractal}.

\subsection{Localization properties}
\label{app:VcEc}

The localization properties of the single-particle states are found by solving the Hamiltonian~(1) of the main paper with $g=0$ using exact diagonalization, and computing the inverse participation ratio (IPR),
\begin{equation}\label{eq:IPR}
\IPR_n = \frac{\int dx\, \vert\Psi_n(x)\vert^{4}}{\left(\int dx\, \vert\Psi_n(x)\vert^{2} \right)^2},
\end{equation}
where $\Psi_n$ is the $n$-th eigenstate.
The IPR scales as $\IPR_n\sim1/L$ for the extended state and as $\IPR_n\sim1$ for a localized state.

 \begin{figure*}[b!]
\centering
     \begin{subfigure}{
         \centering
         \includegraphics[height = 0.16\textheight]{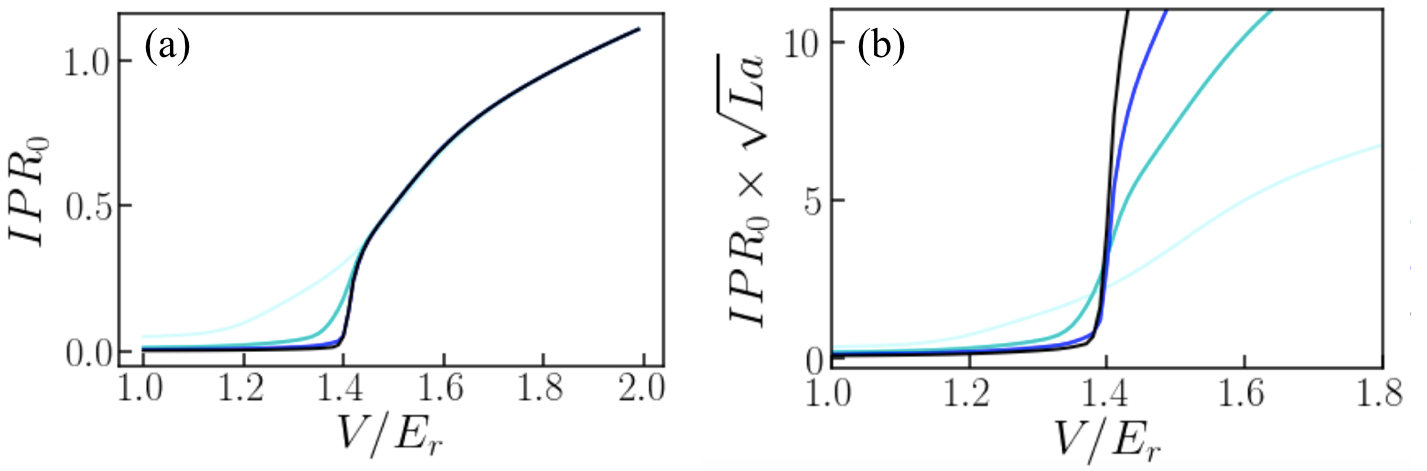}
         \label{fig:1-1}}
     \end{subfigure}
     \begin{subfigure}{
         \centering
        \includegraphics[height =  0.16\textheight]{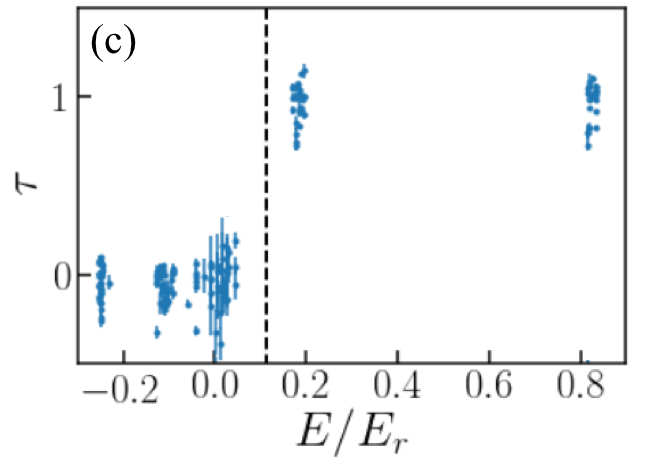}
         \label{fig:1-2}}
     \end{subfigure}
\caption{\label{fig:vc}
Critical potential and mobility edge for the single-particle problem in the bichromatic lattice with $r \simeq 0.807$.
(a)~Ground-state IPR versus the quasiperiodic amplitude $V$ for various system sizes.
Darker lines correspond to increasing system sizes, $L/a=50$~(light blue), $200$~(blue), $500$~(dark blue), and $1000$~(black).
(b)~Rescaled IPR of the ground state, $\IPR_0\times\sqrt{La}$ using the same data as in panel~(a).
(c)~Scaling exponent $\tau$ of the IPR as a function of eigenenergy $E$ for the quasiperiodic amplitude $V=1.5\Er$. It is computed for a system size varying from $L/a=200$ to $2000$.
}
 \end{figure*}

\subsubsection{Critical potential}

To determine the critical amplitude $\Vc$ for localization in the bichromatic lattice, we compute the IPR of the ground state ($n=0$) versus the quasiperiodic amplitude $V$ for various system lengths $L$, see Fig.~\ref{fig:vc}(a). Increasing the potential $V$, we find a clear transition between an extended phase where the IPR is vanishingly small and a localized phase where the IPR is finite.
The transition gets sharper and sharper when the system size $L$ increases [corresponding to darker blue lines on Fig.~\ref{fig:vc}(a)]. Since the rescaled IPR scales as $\IPR_0\times\sqrt{La}\sim1/\sqrt{L}$ in the extended phase and as $\IPR_0\times\sqrt{La}\sim\sqrt{L}$ in the localized phase, an accurate value of the critical potential may be found
by plotting this quantity versus $V$ for various system lengths $L$. The critical potential is then the crossing point of these curves. It yields $\Vc/\Er\simeq1.375\pm0.008$ for $r \simeq 0.807$, see Fig.~\ref{fig:vc} (b).

\subsubsection{Mobility edge}
To determine the ME, 
we compute the IPR as a function of eigenenergy $E$ for a fixed value of the potential amplitude $V$ and various system sizes of the system $L$. In all cases, we find the scaling $\IPR \sim L^{-\tau}$, and always get either $\tau\simeq1$, corresponding to an extended state, or $\tau\simeq0$, corresponding to a localized state.
Increasing the energy $E$ for a fixed value of the potential amplitude above the critical point, $V>\Vc$, we find that the scaling exponent $\tau$ abruptly jumps from $\tau\simeq0$ to $\tau\simeq1$, see Fig.~\ref{fig:vc}(c). The transition between these two values is the ME (black dashed line). More precisely, the ME is always found in a gap and we define $\Ec$ as the energy at the center of this gap as in Ref.~\cite{yao2019}.
For $V=1.5\Er$, as considered in Fig.~\ref{fig:vc}(c), we find $\Ec \simeq 0.115\Er$. It corresponds to the blue dashed line on Fig.~1(a2) of the main paper.
For $V=2\Er$, corresponding to Fig.~1(a3) of the main paper, a similar analysis yields $\Ec \simeq 1.2\Er$, which is beyond the range plotted in this figure.
The case $V=\Er$, corresponding to Fig.~1(a1) of the main paper, is below the critical potential and there is no ME.

\subsection{Fractality of the single-particle spectrum and relation to that of the Mott lobes}
\label{app:fractal}
\begin{figure}[t!]
         \centering
         \includegraphics[width = 0.65\columnwidth]{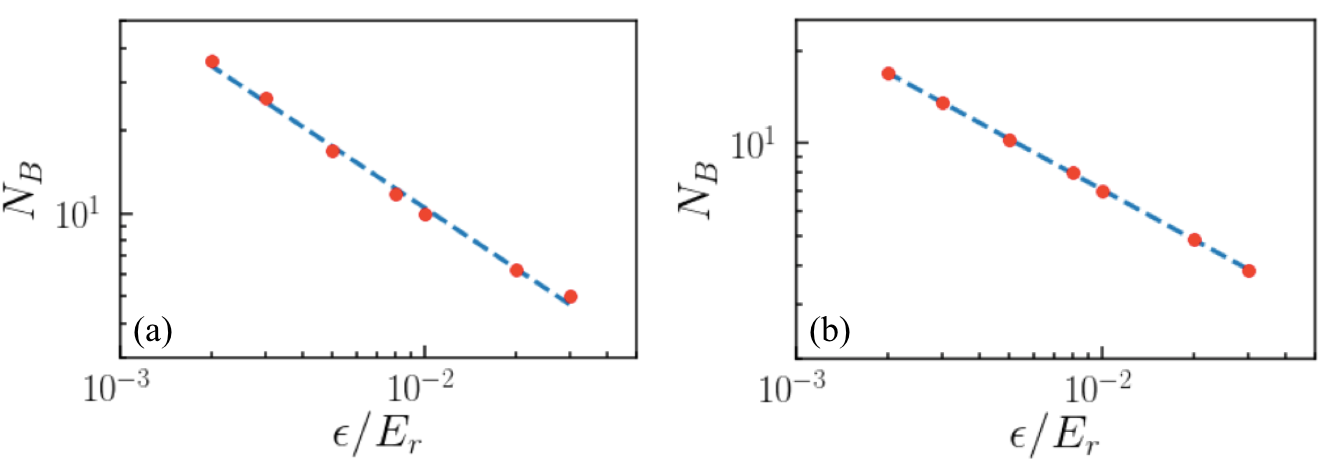}
\caption{\label{dh-cal}
Fractal behaviour of the energy spectrum for single particles in the shallow bichromatic lattice.
Shown are plots of the energy-box counting number $\NB$ as a function of energy resolution $\epsilon$, for two amplitudes of the bichromatic potential:
(a)~$V=1\Er$ and (b)~$V=1.5\Er$.
In both cases, the considered energy ranges $[E_1,E_2]$ are the same as those of the chemical potential $\mu$ on Fig.~1 of the main paper.
}
 \end{figure}

Here, we recall the definitions of the box counting number and the associated Hausdorff dimension of the single-particle spectrum, see Ref.~\cite{yao2019} for further details.
The energy-box counting number within the energy range $[E_1,E_2]$ is the quantity
\begin{equation}\label{nbox}
\NB(\epsilon)=\lim_{q\rightarrow0_+}\int_{E_1}^{E_2}\frac{dE}{\epsilon}\  \big[n_\epsilon(E)\big]^q,
\end{equation}
where $n_\epsilon (E)$ is the integrated density of states (IDOS) per unit lattice spacing,
\ie\ the number of energy eigenstates within the energy range $[E-\epsilon/2,E+\epsilon/2]$, divided by $L/a$.
The quantity $n_\epsilon (E)/\epsilon$ may be interpreted as the density of states per unit lattice spacing for an energy resolution $\epsilon$.
In the limit $q \rightarrow 0_+$, the quantity $\big[n_\epsilon(E)\big]^q$ approaches $1$ if $n_\epsilon(E) \neq 0$
and $0$ if $n_\epsilon(E) = 0$. It thus gives the number of $\epsilon$-wide boxes needed to cover the spectrum in the energy range $[E_1,E_2]$. For a fractal spectrum, it scales as $\NB\sim\epsilon^{-\DH}$, where $\DH$ the spectral Hausdorff dimension. In Fig.~\ref{dh-cal}, we plot $N_B$ versus $\epsilon$ for two values of the quasiperiodic amplitude, namely $V=\Er$ and $V=1.5\Er$. The energy ranges considered here are the same as those of the chemical potential $\mu$ on Fig.~1 of the main paper. For both amplitudes of the quasiperiodic potential (and all cases considered in this work), we find a linear behavior in log-log scale, compatible with the scaling law, $\NB\sim\epsilon^{-\DH}$, which is the characteristic of a fractal spectrum. Fitting the latter to the numerical data, we find $\DH=0.74\pm0.03$ for $V=\Er$ and $\DH=0.54\pm0.01$ for $V=1.5\Er$.

As discussed in the main paper, the mapping of free fermions onto hard-core bosons, relevant in the Tonks-Girardeau limit ($\aOneD \rightarrow 0$) amounts to substitute the energy to the chemical potential ($E \rightarrow \mu$), the IDOS to the compressibility ($n_\epsilon \rightarrow \kappa$), and the energy resolution to the temperature times the Boltzmann constant ($\epsilon \rightarrow \kB T$). Then, comparing Eq.~(\ref{nbox}) to Eq.~(3) of the main paper, we find that, up to the factor $\kB T / (\mu_2-\mu_1)$, the single-particle box counting number $\NB$ maps onto the compressible phase fraction of hard-core bosons $\mathcal{K}$:
\begin{equation}\label{eq:app:box-counting}
\mathcal{K} \sim \frac{\kB T}{\mu_2-\mu_1} N_\textrm{\tiny B}(\epsilon=\kB T),
\end{equation}
which is Eq.~(5) of the main paper.

\subsection{Smoothing of the spectral gaps at finite temperature}
As discussed in the main paper, Bose-Fermi mapping in the hard-core limit allows us to write the equation of state
$$\rho(\mu)\simeq \frac{1}{L}\sum_j \fED(E_j-\mu)
\qquad \textrm{with} \qquad
\fED(E)=\frac{1}{e^{E/\kB T}+1}$$
the Fermi-Dirac distribution and $E_j$ the energy of the $j$-th eigenstate of the single-particle Hamiltonian.
Owing to the fractality of the single-particle spectrum~\cite{yao2019}, at strictly zero temperature, $\rho(\mu)$ is a discontnuous step-like function at any scale.

\begin{figure}[t!]
         \centering
         \includegraphics[width = 0.8\columnwidth]{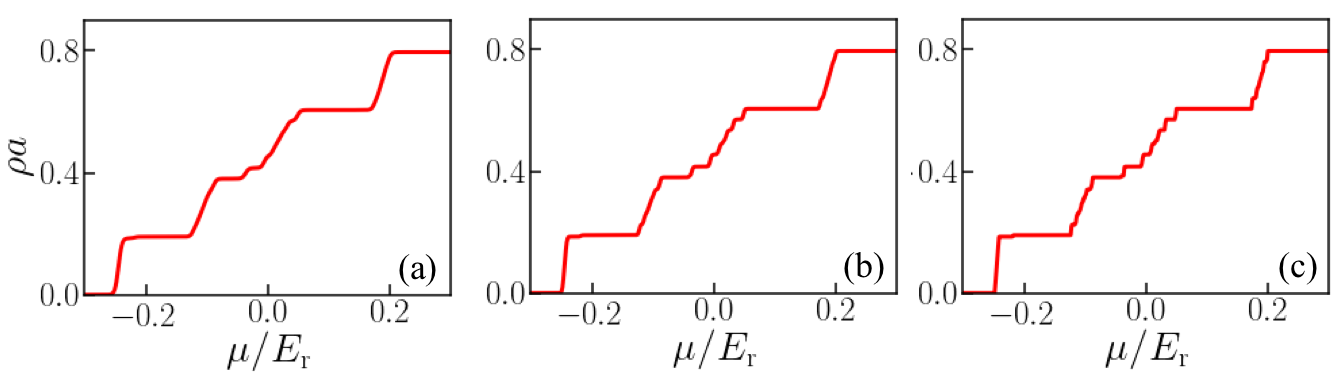}
\caption{\label{fig-sup-FD}
Equation of state $\rho(\mu)$ for free fermions, equivalent to hard-core bosons.
(a)~Same as Fig.~1(a2) in the main paper,
$V=1.5\Er$, $r=0.807$, $T=2\times 10^{-3}\Er/\kB$.
(b) and (c) are same as panel~(a) but with smaller temperatures, $T=8\times 10^{-4}\Er/\kB$ and $T=1.6\times 10^{-4}\Er/\kB$, respectively.
}
 \end{figure}

Any finite temperature $T$ smoothes out all the gaps smaller than the typical energy scale $\kB T$. For instance, Fig.~\ref{fig-sup-FD}(a) reproduces the equation of state plotted on the left-hand side of the Fig.~1(a2) of the main manuscript, corresponding to the temperature $T=2\times 10^{-3}\Er/\kB$. One sees compressible regions between the plateaus. When the temperature decreases, however, new plateaus appear within these compressible regions. See for instance Fig.~\ref{fig-sup-FD}(b) and (c), which correspond to the temperatures $T=8\times 10^{-4}\Er/\kB$ and $T=1.6\times 10^{-4}\Er/\kB$.
This is consistent with the expected pure step-like equation of state expected at strictly zero temperature.
The new plateaus observed on Fig.~\ref{fig-sup-FD}(b) and (c) correspond to Mott gaps for interacting bosons. They, however, appear at a stronger interaction strength than that considered in the manuscript. They are thus irrelevant to our discussion.

\section{Quantum Monte Carlo calculations}
\label{sec:ManyBody}

In this section, we briefly introduce the quantum Monte Carlo approach we use throughout the main paper and present typical results for the particle density $\rho$, the compressibility $\kappa$, the superfluid fraction $\fs$, and the one-body correlation function $g_1(x)$. All our calculations are performed for a system size $L=83a$, where $a$ is the lattice spacing of the first lattice, and for the bichromatic potential in Eq.~(2) of the main paper with the lattice spacing ratio $r \simeq 0.807$.

\subsection{Path integral quantum Monte Carlo algorithm}

Accurate values of the compressibility and the superfluid fraction are found using large-scale, path-integral quantum Monte Carlo (QMC) calculations in continuous space, within the grand-canonical ensemble at the chemical potential $\mu$ and the temperature $T$.
It yields accurate estimates of the thermodynamic averages of any observable $A$,
\begin{equation}\label{eq:app:avA}
\langle A \rangle = \frac{\textrm{Tr}\left[\e^{-\beta(\mathcal{H}-\mu N)} A \right]}{\textrm{Tr}\left[\e^{-\beta(\mathcal{H}-\mu N)}\right]},
\end{equation}
where $\mathcal{H}$ is the Hamiltonian, $N$ the number of particles operator, $\beta=1/\kB T$, and $\textrm{Tr}$ the trace operator.
The worm algorithm~\cite{boninsegni2006,Boninsegni2006b} spans a large number of boson configurations within both the physical Z-sector (closed worldlines) and an unphysical G-sector (worms, \ie\ worldlines with open ends).
The average number of particles $N$ is found from the statistics of worldlines within the Z-sector. It yields the particle density $\rho=N/L$, where $L$ is the system size, and the compressibility $\kappa=\partial \rho/\partial \mu$.
The superfluid fraction $\fs=\ups/\rho$ is found from the superfluid stiffness $\ups$, computed using the winding number estimator~\cite{ceperley1995}.
The one-body correlation function $g_1(x)=\int \frac{dx^{\prime}}{L}\langle \Psi(x^{\prime}+x)^\dagger\Psi(x^{\prime})\rangle$ is found from the statistics of worms with open ends at $x'$ and $x'+x$ within the G sector~\cite{boninsegni2006,Boninsegni2006b}.

Details about our implementation of QMC algorithm are discussed in previous work~\cite{carleo2013,boeris2016,yao2018}.

\subsection{Particle density $\rho$, compressibility $\kappa$, and superfluid fraction $\fs$}

In the main paper, the zero-temperature phases (superfluid, SF; Mott insulator, MI; Bose glass, BG) are identified via the values of the compressibility $\kappa=\partial\rho/\partial\mu$ and the superfluid fraction $\fs=\ups/\rho$,
where $\rho=N/L$ is the particle density, $\mu$ the chemical potential, and $\ups$ the superfluid stiffness.
In addition, we compute the one-body correlation function $g_1(x)$ and, for insulating phases, the correlation length $\xi$, such that $g_1(x) \sim \exp\left(-\vert x\vert/\xi\right)$. It allows us to distinguish the BG regime from the normal fluid (NF) regime at finite temperature, see Table~\ref{table:judge}.

\vspace{-0.2cm}

\begin{table}[h!]
  \centering
  \begin{tabular}{|c|c|c|c|}
    \hline
    Phase  & Superfluid fraction $f_s$ & Compressibility $\kappa$ & $T$-dep.\ of corr.\ length $\partial \xi/ \partial T$ \\ \hline
    Superfluid (SF)  & $\neq 0$ & $\neq 0$ & $/$ \\ \hline
    Mott-insulator (MI)  & $= 0$ & $= 0$ & $\sim 0$ \\ \hline
    Bose-glass (BG)  & $= 0$ & $\neq 0$ & $\sim 0$ \\ \hline
    Normal fluid (NF)  & $= 0$ & $\neq 0$ & $\neq 0$ \\ \hline
  \end{tabular}
  \caption{\label{table:judge}
  Identification of the (zero-temperature) quantum phases and finite-temperature regimes from quantum Monte Carlo calculations. Note that the one-body correlation function $g_1(x)$ is algebraic in the superfluid regime and the correlation length $\xi(T)$ is not defined.}
\end{table}

\vspace{-0.2cm}

\begin{figure}[t!]
         \centering
         \includegraphics[width = 1.0\columnwidth]{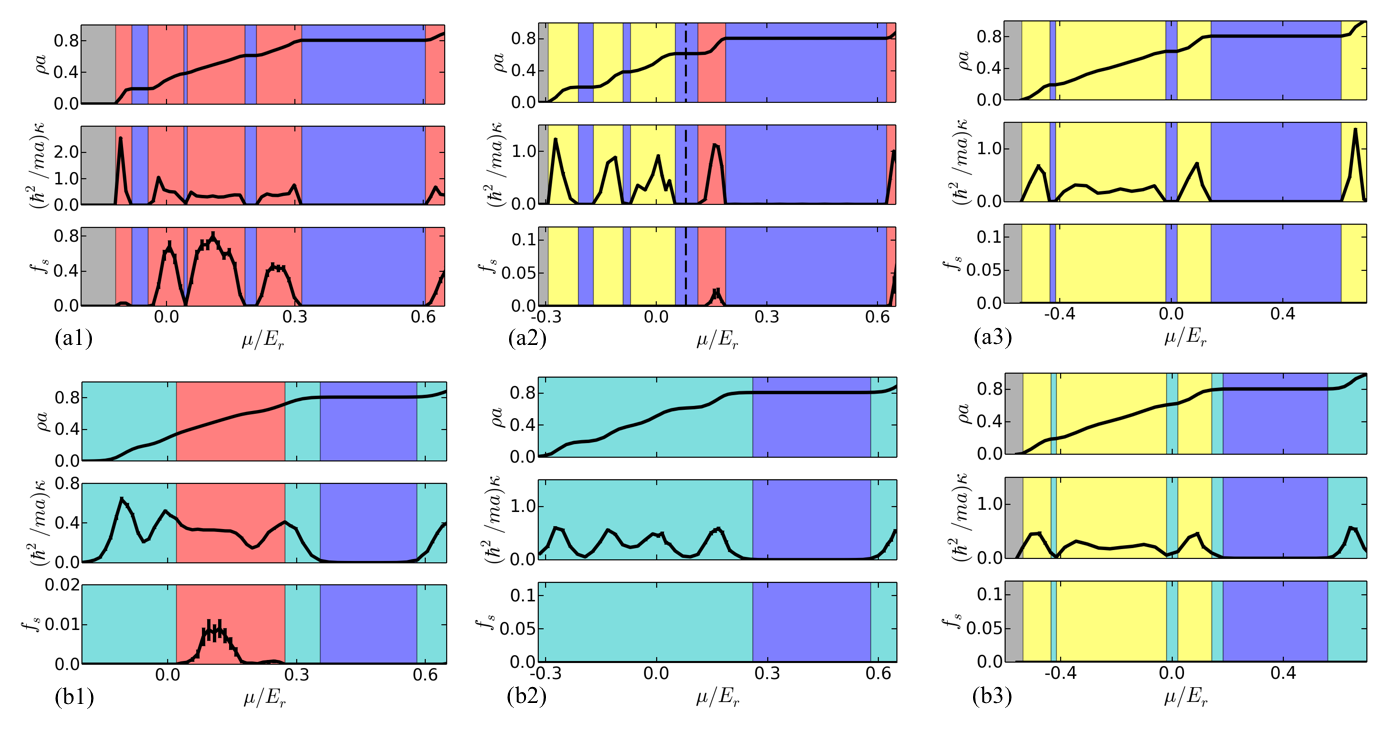}
\caption{\label{fig:qmc-supple}
Typical QMC results for the density $\rho$, the superfluid fraction $\fs$, and the compressibility $\kappa$ as a function of the chemical potential $\mu$. The various panels are cuts of the diagrams of Fig.~1 of the main paper at the interaction strength $-\aOneD/a=0.1$:
(a1)~$V=\Er$ and $\kB T/\Er=0.001$,
(a2)~$V=1.5\Er$ and $\kB T/\Er=0.002$
(a3)~$V=2\Er$ and $\kB T/\Er=0.002$,
(b1)-(b3)~Same as in panle~(a) but at temperature $\kB T/\Er=0.015$.
\vspace{-0.2cm}
}

 \end{figure}

Figure~\ref{fig:qmc-supple} shows typical results for the particle density $\rho$, the compressibility $\kappa$, and the superfluid fraction $\fs$. The six panels correspond to cuts of the six diagrams of Fig.~1 of the main paper at the interaction strength $-\aOneD/a=0.1$.

\vspace{0.25cm}
\noindent\textit{Figure~\ref{fig:qmc-supple}(a1)~[$V=\Er$; $T=0$]:}
We find an alternation of compressible ($\kappa>0$) and incompressible ($\kappa=0$) phases, in exact correspondance with superfluidity: the compressible phases always have a finite superfluid fraction ($\fs>0$) while the incompressible phases are always non-superfluid ($\fs=0$). They correspond to SF~(red areas) and MI phases~(blue areas), respectively. The absence of a BG phase is consistent with the fact that the potential amplitude is below the critical potential, $V<\Vc \simeq 1.38\Er$.

\vspace{0.25cm}
\noindent\textit{Figure~\ref{fig:qmc-supple}(a2)~[$V=1.5\Er$; $T=0$]:}
Here we find a similar behaviour for large enough chemical potential, $\mu \gtrsim 0.1\Er$. For smaller chemical potentials, however, we find clear signatures of BG phases, corresponding to a compressible insulator ($\kappa>0$ and $\fs=0$, yellow areas).
Here, the BG phases are separated by MI phases ($\kappa=0$ and $\fs=0$, blue areas).
As expected, in the strongly-interacting limit, the BG phase appear only for $\mu<\Ec \simeq 0.115\Er$, \ie\ the single-particle mobility edge~(dashed black line).

\vspace{0.25cm}
\noindent\textit{Figure~\ref{fig:qmc-supple}(a3)~[$V=2\Er$; $T=0$]:}
In this case, the Bose gas is non-superfluid, $\fs=0$, in the whole range of the chemical potential considered here. It, however, shows an alternance of compressible  and incompressible phases, corresponding to BG~(yellow areas) and MI~(blue areas) phases, respectively. Note that for $V=2\Er$ the single-particle mobility edge is $\Ec \simeq 1.2\Er$, which is beyond the considered range of the chemical potential.

\vspace{0.25cm}
\noindent\textit{Figure~\ref{fig:qmc-supple}(b1-b3)~[$T=0.015\Er$]:}
The lower panel shows the finite-temperature counterpart of the upper panel. The various regimes are characterized by the same criteria as for zero temperature.
First, we find regimes with vanishingly small compressibility and superfluid fraction. They correspond to regimes where the zero-temperature MI is unaffected by the finite temperature effects (blue areas, all panels).
Second, although superfluidity is absent in the thermodynamic limit, we find compressible regimes with a clear non-zero superfluid fraction in our system of size $L=83a$ for weak enough quasiperiodic potential~(red areas, left panel).
We refer to such regimes as finite-size superfluids. We have checked that the one-body correlation function is, consistently, algebraic over the full system size in these regimes (see below).
Third, we find insulating, compressible regimes~($\fs=0$ and $\kappa>0$). At finite temperature, however, the values of $\fs$ and $\kappa$ are not sufficient to distinguish the BG pand NF phases, which are thus discriminated via the temperature dependence of the correlation length $\xi(T)$: the absence of temperature dependence shows that the quantum phase is unaffected by the thermal fluctuations and the corresponding regimes are identified as the BG~(yellow areas). Conversely, the regimes where the correlation length shows a sizable temperature dependence are identified as the NF~(light blue areas).

\subsection{One-body correlation function}
Here we discuss the behaviour of the one-body correlation function in various regimes. It reads as
\begin{equation}\label{nbox}
g_1(x) = \int \frac{dx^{\prime}}{L}\langle \Psi(x^{\prime}+x)^\dagger\Psi(x^{\prime})\rangle,
\end{equation}
where $\Psi(x)$ is the Bose field operator. In the QMC calculations, it is computed from the statistics of worms  with open ends at $x'$ and $x'+x$ within the G-sector~\cite{boninsegni2006,Boninsegni2006b}.

 \begin{figure}[h!]
         \centering
         \includegraphics[width = 0.62\columnwidth]{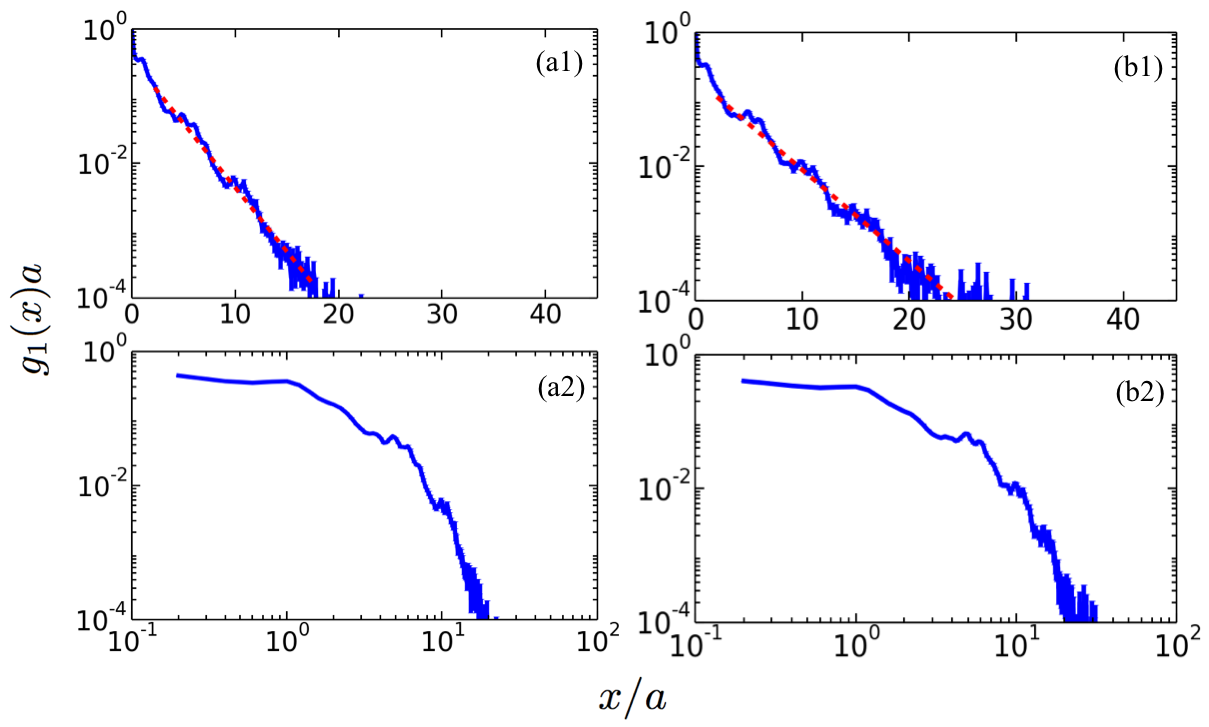}
\caption{\label{fig:g1-supple-1}
One-body correlation function $g_1(x)$ for the same parameters as in Fig.~2(a) of the main paper for two different temperatures: (a)~$T=0.004\Er/\kB$~(BG regime) and (b)~$T=0.04\Er/\kB$~(NF regime).
The upper and lower panels show plots of the same data in semi-log and log-log scales, respectively.
The dashed red lines indicate linear fits to $g_1(x)$ in semi-log scale. It yields the coherence lengths
(a)~$\xi \simeq 2.86a$ and (b)~$\xi=1.81a$, respectively.
}
 \end{figure}
 \begin{figure}[h!]
         \centering
         \includegraphics[width = 0.62\columnwidth]{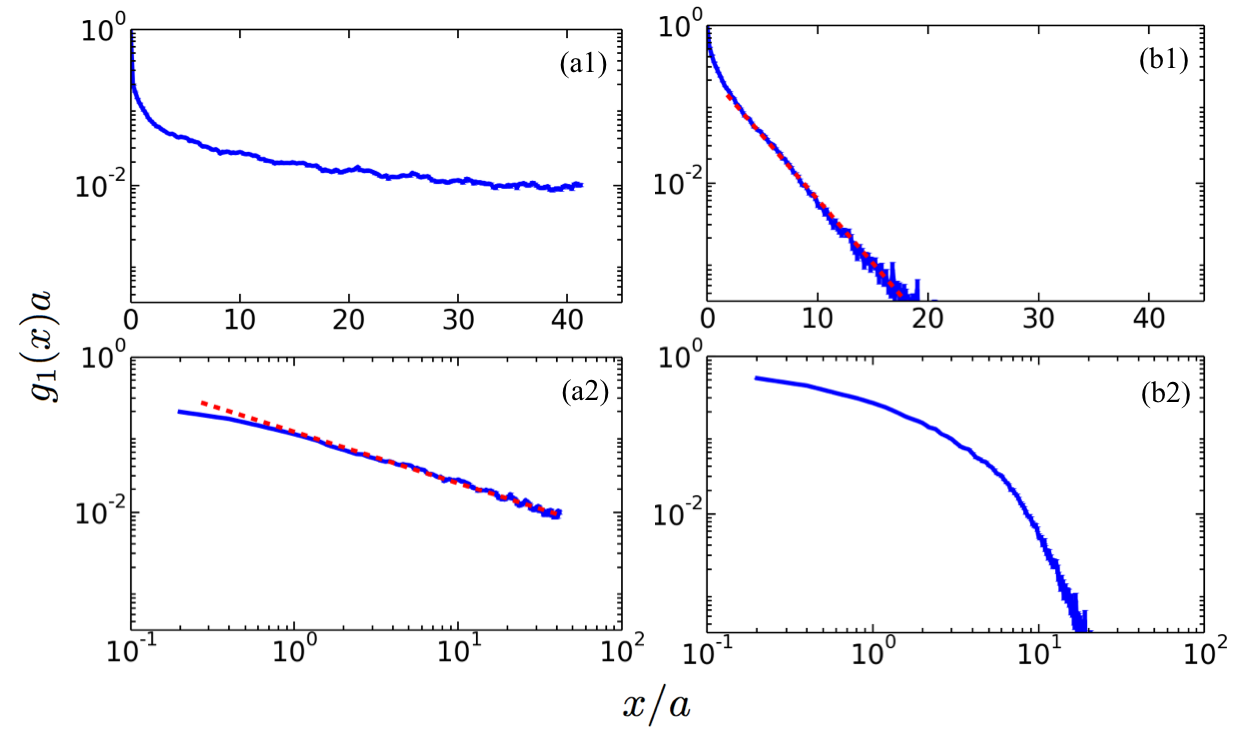}
\caption{\label{fig:g1-supple-2}
One-body correlation function $g_1(x)$ for the same parameters as in Fig.~2(b) of the main paper for two different temperatures: (a)~$T=2.4\times10^{-3}\Er/\kB$~(finite-size SF regime) and (b)~$T=0.03\Er/\kB$~(NF regime).
The upper and lower panels show plots of the same data in semi-log and log-log scales, respectively.
The dashed red lines indicate linear fits to $g_1(x)$ in semi-log scale or log-log scale.
}
 \end{figure}


As discussed above, the one-body correlation function is mainly used to discriminate the BG regime from the NF regime at finite temperature, which are both compressible insulators.
Figure~\ref{fig:g1-supple-1} shows the behaviour of $g_1(x)$ for parameters as in Fig.~2(a) of the main paper and two different temperatures. The upper and lower panels show plots of the same data in semi-log and log-log scales, respectively. We find, in both cases, that $g_1(x)$ is better fitted by an exponential function, $g_1(x) \sim \exp(-\vert x \vert/\xi)$, rather than an algebraic function. This is consistent with the expected behaviour in insulating regimes.
Fitting the linear slope in semi-log scale (dotted red line), we extract the correlation length $\xi(T)$.
The values of $\xi(T)$ plotted on Fig.~2(a) of the main paper are all found from the similar way of fits.

Figure~\ref{fig:g1-supple-2} shows the counterpart of the previous plots for the parameters of Fig.~2(b) of the main paper and two different temperatures.
On the one hand, the left panel corresponds to the temperature $ T=2.4\times10^{-3}\Er/\kB$, where we find a finite-size SF [see Fig.~2(b) of the main paper]. Consistently, the one-body correlation function is well fitted by an algebraic function~(dotted red line), but not by an exponential function, over the full system size.
On the other hand, the right panel corresponds to the temperature $ T=3\times10^{-2}\Er/\kB$, where we find a compressible insulator with a temperature-dependent correlation length [NF, see Fig.~2(b) of the main paper]. Consistently, the one-body correlation function is here better fitted by an exponential function~(dotted red line) than by an algebraic function, over the full system size.

 \begin{figure}[h!]
         \centering
         \includegraphics[width = 0.95\columnwidth]{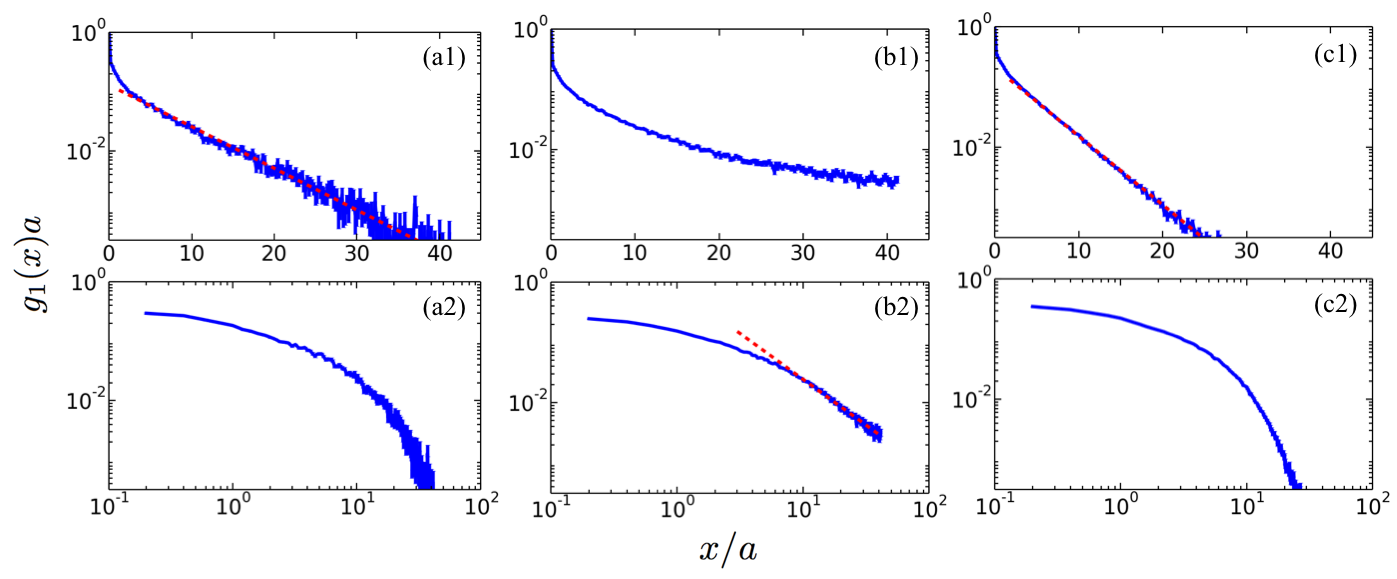}
\caption{\label{fig:g1-supple-3}
One-body correlation function $g_1(x)$ for the same parameters as in Fig.~2(d) of the main paper for three different temperatures:
(a)~$T=8\times10^{-4}\Er/\kB$ (MI regime),
(b)~$T=4\times10^{-3}\Er/\kB$ (finite-size SF regime),
and (c)~$T=3\times10^{-2}\Er/\kB$ (NF regime).
The upper and lower panels show plots of the same data in semi-log and log-log scales, respectively.
The dashed red lines indicate linear fits to $g_1(x)$ in semi-log or log-log scale.
}
 \end{figure}

\begin{figure}[h!]
         \centering
         \includegraphics[width = 0.5\columnwidth]{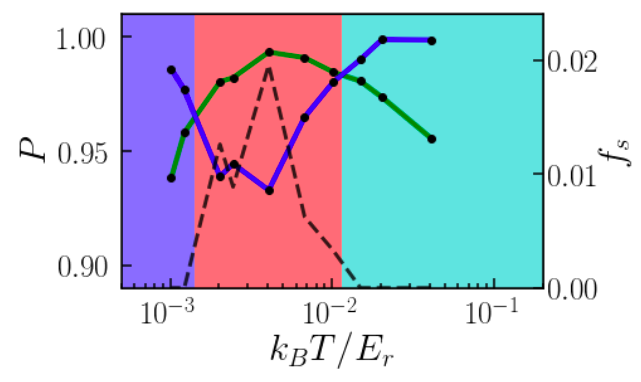}
\caption{\label{bcorrelation}
Pearson's correlation coefficient for linear fits of one-body correlation function $g_1(x)$ for the same parameters as in Fig.~2(d) in semi-log scale (solid blue line) and log-log scale (solid green line).
The colored areas correspond to the MI (dark blue), SF (red), and NF (light blue) regimes, determined from the zero or non-zero value of the superfluid fraction, as in Fig.~2(d).
The black dashed line is the superfluid fraction $\fs$ [reproduced from Fig.~2(d)].
}
 \end{figure}

Finally, Fig.~\ref{fig:g1-supple-3} shows the $g_1(x)$ functions, in both semi-log~(upper panel) and log-log~(lower panel) scales, for the parameters of Fig.~2(d) of the main paper and three different temperatures:
(a)~$T=8\times10^{-4}\Er/\kB$, corresponding to the MI regime,
(b)~$T=4\times10^{-3}\Er/\kB$, corresponding to the finite-size SF regime,
and (c)~$T=3\times10^{-2}\Er/\kB$, corresponding to the NF regime [see Fig.~2(d) of the main paper].
Consistently, we find that $g_1(x)$ is better fitted by an exponential function in the insulating regimes~[MI and NF, panels~(a) and~(c)] and by an algebraic function in the finite-size SF regime [panel~(b)].
This is further confirmed by the calculation of the Pearson correlation coefficients $P$
for a linear fit of $g_1(x)$ in semi-log and log-log scales, see Fig.~\ref{bcorrelation}.
The closer $P$ is to unity, the better the linear fit.
Figure~\ref{bcorrelation} confirms that the correlation function is closer to an exponential function in the MI (dark blue) and NF (light blue) regimes, and closer to an algebraic function in the superfluid regime (red).
Note that in Fig.~\ref{bcorrelation}, the colored areas are determined according to the zero or non-zero value of the superfluid fraction $\fs$ (reproduced as the black dashed line from the data of the Fig.~2(d) of the main paper). The turning points match with the crossings of the Pearson curves corresponding to algebraic and exponentials fits, respectively.

\section{Bose glass at higher temperatures}
\label{sec:exp}

\begin{figure}[h!]
         \centering
         \includegraphics[width = 0.9\columnwidth]{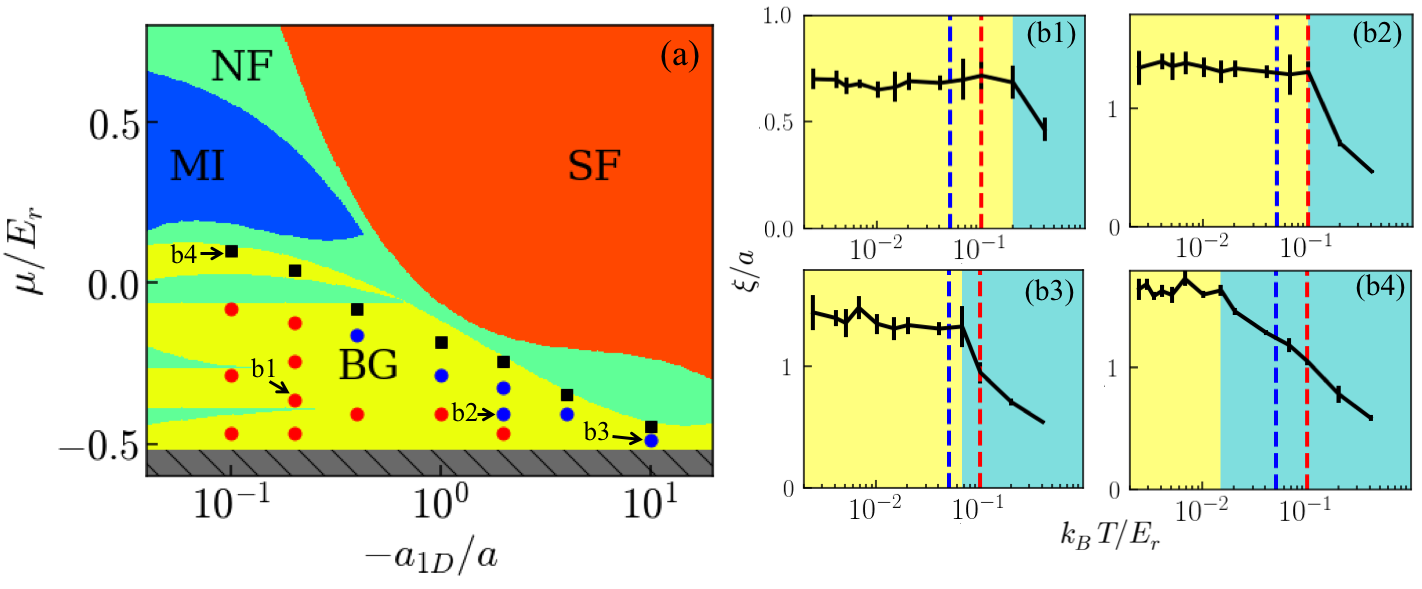}
\caption{\label{experiment-check}
Existence of the BG regime for $V=2\Er$ at higher temperatures.
(a)~Reproduction of the Fig.~1(b3) of the main paper, indicating the MI (blue), SF (red), BG (yellow), and NF (green) regimes at the temperature $T=0.05\Er/\kB$.
The markers indicate points where the melting temperature $T^*$ of the BG phase
is
$\kB T^* / \Er < 0.05$~(black squares),
$0.05 < \kB T^* / \Er < 0.1$~(blue disks)
or $\kB T^* / \Er > 0.1$~(red disks).
(b)~Temperature dependence of the correlation length $\xi$ for the four points indicated on panel~(a).
The dashed blue and red lines indicate the temperatures $T=0.05\Er/\kB$ and  $T=0.1\Er/\kB$, respectively.
The background colors indicate the BG (yellow) and NF (light blue) regimes. 
}
 \end{figure}

In the main paper, the diagrams of the lower row in Fig.~1 are computed at the finite temperature $T=0.015\Er/\kB$. It corresponds to the temperature $T=1.5$nK for 1D $^{133}$Cs atoms, corresponding approximately to the lowest temperatures reported in Ref.~\cite{meinert2015}. For the amplitude of the quasiperiodic potential $V=2\Er$, its shows a sizable BG regime, see Fig.~\ref{experiment-check}(a), which reproduces the Fig.~1(b3) of the main paper. The BG should be observable in such an experiment.

Moreover, we have checked that a sizable BG regime survives up to higher temperatures, which can be achieved in 1D ultracold gases. For instance, the experiment of Ref.~\cite{meinert2015} reported temperatures in the range $1$nK$ \lesssim T \lesssim 10$nK  for $^{133}$Cs atoms, corresponding to $0.015 \lesssim\kB T / \Er \lesssim 0.15$.
The experiment of Ref.~\cite{derrico2014} was operated at $T \simeq 15$nK  for $^{39}$K atoms, corresponding to $\kB T / \Er \simeq 0.07$.
We have studied the temperature dependence of the correlation length $\xi$ as determined from exponential fits to the one-body correlation function $g_1(x)$. Some examples are shown on Fig.~2 of the main paper as well as in Figs.~\ref{experiment-check}(b1)-(b4) of this supplemental material. The melting temperature $T^*$ of the BG is the temperature where $\xi(T)$ starts to decrease.
The points in the BG regime we have check are indicated by markers on Fig.~\ref{experiment-check}(a):
black squares correspond to cases where
$\kB T^* / \Er < 0.05$~[see for instance Fig.\ref{experiment-check}(b4)],
blue disks to cases where $0.05 < \kB T^* / \Er < 0.1$~[see for instance Fig.~\ref{experiment-check}(b3)],
and red disks to cases where $\kB T^* / \Er > 0.1$~[see for instance Fig.~\ref{experiment-check}(b1) and (b2)].
It shows that sizable BG regimes are still found at $T \simeq 0.05\Er/\kB$ and at $T \simeq 0.1\Er/\kB$, and should thus be observable in experiments such as those of Refs.~\cite{derrico2014,meinert2015}.

\begin{thebibliography}{10}
\bibitem{note:size}
{For the finite system size $L=83a$, the ratio $r\simeq0.807 \simeq 67/83$ is a
  fair approximation for an irrational number since no periodicity is
  realized.}
\bibitem{note:Tstar}
{The gap of the smallest lobes does not vary much with the interaction strength
  while that of the largest lobes does; see Fig.~\ref{fig:phase}. It explains
  that only $T_2^{\star}$ varies significantly with the interaction strength in
  Fig.~\ref{fig:fractalMIlobes}.}
\end{thebibliography}
\end{document}